\documentclass[pra,reprint,fleqn]{revtex4-2}
\usepackage[english]{babel}
\usepackage{graphicx}
\usepackage{amsmath}
\usepackage[utf8]{inputenc}
\usepackage{xcolor}
\usepackage{physics}
\usepackage{mathtools}
\usepackage{bm}
\usepackage[normalem]{ulem}
\usepackage{mathtools}
\renewcommand{\i}{\mathrm{i}}
\usepackage[charter,expert]{mathdesign}
\usepackage[scaled=0.96]{XCharter}

\def\eps{\varepsilon}
\def\d{\mathrm{d}}
\def\i{\mathrm{i}}
\def\e#1{\mathrm{e}^{{#1}}}
\def\vec#1{\boldsymbol{#1}}

\begin{document}

\title{Zero-energy photoelectric effect}
\author{Sajad Azizi, Ulf Saalmann and Jan M. Rost}
\affiliation{Max-Planck-Institut f{\"u}r Physik komplexer Systeme,
 N{\"o}thnitzer Str.\ 38, 01187 Dresden, Germany}
\date{\today}

\begin{abstract}\noindent
We predict a near-threshold (``zero energy'') peak in multi-photon ionization for a dynamical regime where the photon frequency is large compared to the binding energy of the electron. 
The peak position does not depend on the laser frequency, but on the binding energy and the pulse duration. 
The effect originates from the fact that bound-continuum dipole transitions are stronger than continuum-continuum ones. 
To clearly observe this zero-energy photoelectric effect, the spectral width of the laser pulse should be comparable to the binding energy of the ionized orbital, and the second ionization potential should be larger than the photon energy. 
This suggests negative ions as ideal candidates for corresponding experiments.
\end{abstract}

\maketitle

\noindent
A threshold for an observable $A$ indicates the transition of the system from one regime to another upon change of the relevant parameter $\eps$ and provides therefore important information
about the system. Often, the observable changes near threshold at $\eps_0$ with a certain power of the parameter, 
i.e., $A (\eps{\to}{+}\eps_0) \propto \left(\eps/\eps_0{-}1\right)^\alpha$. 
Probably best known are thermodynamical variables near phase transitions \cite{st71}, but also quantum critical points in condensed matter \cite{sa00} or fragmentation/ionization thresholds in atomic and molecular physics \cite{ro01} are examples. In the latter case
the so-called Wigner threshold law \cite{wi48} for fragmentation cross sections of particles under short-range forces is a universal property, similar as its counterpart for long-range (Coulomb) forces, the Wannier law \cite{wa53,ro94}. 

The Wigner law has been verified in fragmentation scenarios including multi-photon detachment of a negative ions \cite{rehe01}, which were a popular target for theoretical considerations \cite{krfa+06} regarding above-threshold ionization (ATI). 
However, so far it has gone unnoticed that the combination of the Wigner threshold behavior with \emph{short} intense pulses can give rise
to a peculiar zero-energy photoelectric effect (ZEPE) with a characteristic maximum, whose position at very low photo-electron energy does not depend on the photon energy, but on the duration of the (short) laser pulse.
That short intense pulses can lead to unusual electron dynamics has been pointed out in the context of non-adiabatic photo-ionization \cite{toto+09,nisa+18} where short pulses can even be used for coherent control \cite{azsa+21}.

\begin{figure}[!b]
 \centering
 \includegraphics[width=0.9\columnwidth]{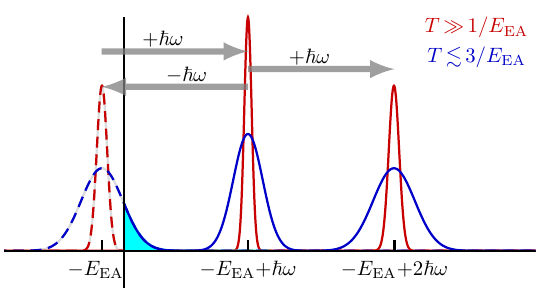}
 \caption{Multi-photon detachment of a an electron, weakly bound by electron affinity $E_{\rm EA}$ to an ion, will give rise to
 peaks in the photo-electron spectrum at energies $E\,{=}\,{-}E_{\rm EA}{+}n\omega$, shown here for $n{=}1,2$. 
 However, following absorption of the 1st photon it is more likely that a 2nd photon is not absorbed but emitted (${-}\hbar\omega$) since the dipole for this transition is much stronger. 
 This process with net-zero energy absorption
 becomes only visible in the spectrum for very short pulses (blue) leading to a broad peak at $-E_{\rm EA}$ whose tail reaches into the continuum (lightblue-shaded), but not for longer pulses (red).}
 \label{fig:sketch}
\end{figure}

The ZEPE effect requires pulses short enough such that their spectral bandwidth $\Delta E$ is larger than the binding energy $E_{\rm EA}$ of the detached electron, typically of the order of a few femtoseconds. This was clearly not the case in the ATI detachment experiments with 800\,nm light.
ZEPE is enabled by a two-photon process,
where a bound electron absorbs a photon and emits a photon, ending
at the same (bound) energy as it started from. Therefore, this mechanism does typically not contribute to the photo-electron spectrum. Only, if the laser pulse is short enough that the spectral peak at the binding energy
``leaks'' into the continuum, see Fig.\,\ref{fig:sketch}, this process becomes visible in form of photo-electrons and can lead, together with the Wigner power law, to a pronounced ZEPE maximum.

Negative ions are an ideal target to clearly identify the ZEPE, since they combine a low electron affinity (EA), i.e., the ionization potential of the negative ion, with a typically large gap to the next ionization potential (IP), i.e., the ionization potential of the (neutral) atom. 
Therefore, a relatively large photon frequency, still fulfilling $E_{\rm EA}<\omega < E_{\rm IP}$, can be chosen, which on the one hand energetically prevents ionization of more deeply bound electrons 
and on the other hand lets the first ATI peak appear at relatively high energy, keeping an energy interval just above threshold pristine for a clean signature of ZEPE.

We will demonstrate the effect in the following with negative ions of hydrogen and oxygen in a simple and transparent fashion. To this end, we use an effective potential
for the bound electron of the negative ion which is designed to reproduce the EA well \cite{gata+80,shta87},
\begin{equation}
 V(r) = - \frac{Z}{r} \frac{b}{1 + c\, [\exp(r/r_0){-}1]}\,,
 \label{eq:potential}
\end{equation}
where $Z$ is the nuclear charge, $Z\,{=}\,1$ and $Z\,{=}\,8$ for hydrogen and oxygen, respectively. We use atomic units unless stated otherwise. 
With parameter values $b\,{=}\,1.1$, $c\,{=}\,1$, and $r_0\,{=}\,0.5292$\,\AA\ the computed $E_{\rm 1s}\,{=}\,{-}0.75$\,eV for H$^{-}$ closely matches the experimental value \cite{lymu+91}. 
Similarly, with $b\,{=}\,1$, $c\,{=}\,1.9607$, and $r_0\,{=}\,0.4689$\,\AA, the computed $E_{\rm 2p}\,{=}\,{-}1.464$\,eV for O$^{-}$ is in good agreement with the experimental value of $-1.461$\,eV \cite{krch+22}. 

With the potential \eqref{eq:potential} the Hamilton operator reads
\begin{equation}
 \label{eq:hamil}
 \widehat{H}(t) = \widehat{\vec{p}}{\,}^2\!/2 + V(r) + \widehat{\vec{p}} \cdot \vec{e}_z \, A_0 g(t) \cos(\omega t),
\end{equation}
where $g(t)\,{=}\,{\exp}({-}t^2/T^2)$ is the laser-pulse envelope with the pulse duration $T\,{=}\,T_{\rm fwhm}/\sqrt{2\ln\!2}$, linearly polarized along the $z$-axis and dipole coupled in velocity gauge. 
In single-active-electron approximation \eqref{eq:hamil}, the propagation of the 3D time-dependent Schr\"odinger equation (TDSE), for processes discussed here, is straightforward, since the number of photons involved is moderate. 
What is challenging, however, is the numerically accurate description of the \emph{very} small detachment probabilities close to threshold. 
To this end we calculate eigenstates and dipole matrix elements for the potentials given in \eqref{eq:potential} for $\ell\,{=}\,0\ldots\ell_{\rm max}$ in a box $r\,{=}\,0\ldots R$ up to a cutoff energy $E_{\rm cut}$. The parameters used are $\ell_{\rm max}\,{=}\,4$, $R\,{=}\,3{\times}10^3$\,\AA\ and $E_{\rm cut}\,{=}\,3$\,keV. 
The TDSE is propagated numerically with this field-free basis.
From the final amplitudes obtained we calculate the photo-electron spectrum. Attaching to every eigenstate a normalized Gaussian, whose height is the absolute square of the amplitude and whose width is given by the level spacing at the corresponding eigenenergy, produces a continuous spectrum. Thereby, the density of states and the detachment process are correctly included.
\begin{figure}[b]
 \centering
 \includegraphics[width=\columnwidth]{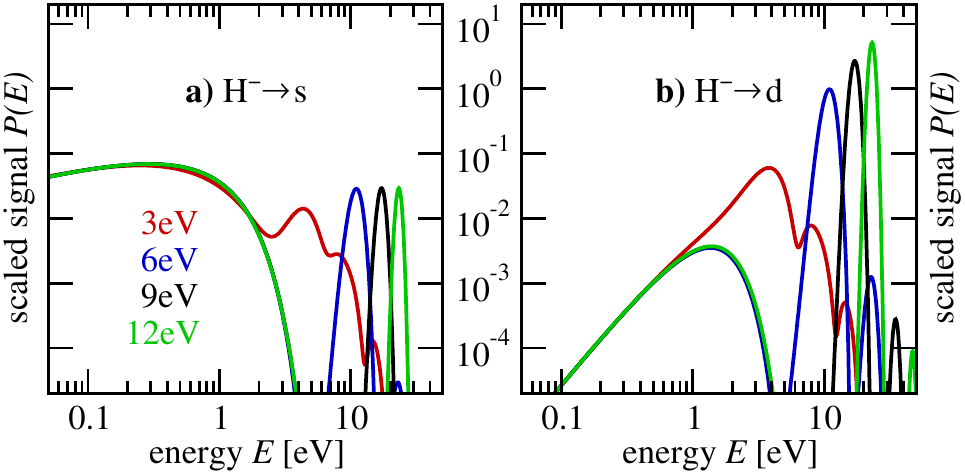}
 \caption{Spectra from H$^-$ for $T_{\rm fwhm}\,{=}\,1$\,fs, $I\,{=}\,10^{14}$W/cm$^2$ and 4 different photon frequencies ($\omega\,{=}\,3,6,9,12$\,eV). Note, while higher-order peaks (at energies $E\,{\gtrsim}\,6$\,eV) shift with increasing $\omega$ to higher energies, the zero-peak position ($E_{\rm s}\,{\approx}\,0.3$\,eV and $E_{\rm d}\,{\approx}\,1.4$\,eV) is independent of $\omega$.
 Spectra are scaled such that they agree for $E\,{=}\,0.1$\,eV.}
 \label{fig:4freqs}
\end{figure}%

Figure\,\ref{fig:4freqs} shows the resulting photo-electron spectra for H$^-$ exposed to a laser pulse of 1\,fs duration and a peak intensity of $I\,{=}\, 10^{14}\,\mathrm{W/cm^2}$ for different photon frequencies. 
The unique feature of the zero-energy photo-effect is evident: There is a pronounced maximum close to threshold ($E_{\rm s}{\approx}0.3$\,eV and $E_{\rm d}{\approx}1.4$\,eV) which is independent of the photon frequency, clearly different in shape from the two-photon ATI peaks which can be identified at the respective energies $E\,{=}\,2\omega\,{-}\,E_{\rm EA}$. 
If the ZEPE maximum is much smaller than and too close to the ATI peaks, typically for higher final partial waves $\ell>0$, then it may get buried under the rise to the two-photon ATI peak, which is the case for H$^{-}{\to}$\,d detachment with 3\,eV photons, as visible in Fig.\,\ref{fig:4freqs}b.

\begin{figure}[t]
 \centering
\includegraphics[width=\columnwidth]{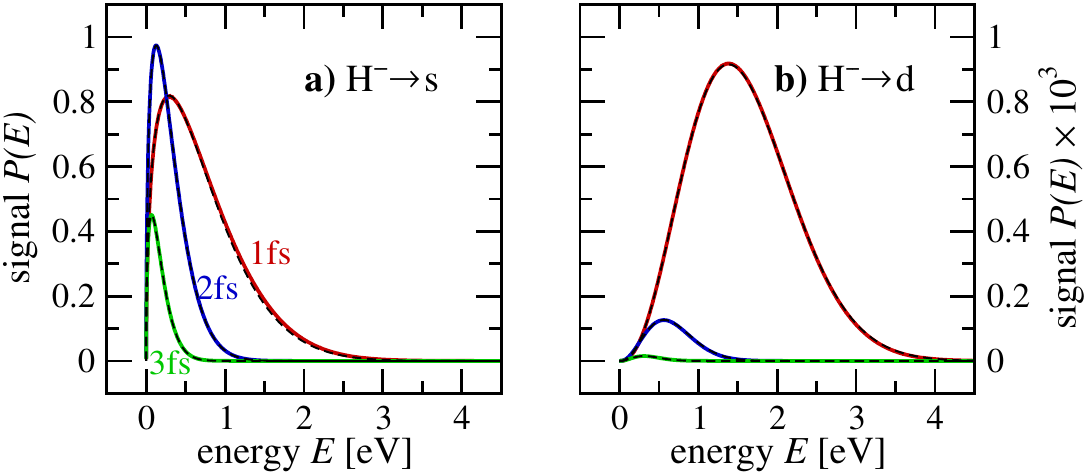}
 \caption{Photo-electron spectrum (colored solid lines) of H$^-$ for a Gaussian pulse with different pulse duration $T_{\rm fwhm} = 1, 2, 3$\,fs, carrier frequency $\omega = 9\,$eV, and intensity $I = 10^{14}\,\mathrm{W/cm^2}$ 
 for the s (a) and the d-channel (b), respectively,
 and fits (black dashed lines) according to Eq.\,\eqref{eq:ZEPE-spectrum}.
 }
 \label{fig:ZEPE-linear}
\end{figure}%
To understand how ZEPE comes about qualitatively and quantitatively, 
we take a closer look at the near-threshold energy range with the spectra of Fig.\,\ref{fig:ZEPE-linear} on a linear scale for different pulse durations $T$. One immediately sees that the spectra depend on $T$, as already anticipated.
The Wigner threshold law \cite{wi48} states that for break-up of two fragments under short-range forces (which is the case for electron detachment), the ionization probability near threshold is given by the available continuum states in momentum space, i.e.,
\begin{equation}
P(E{\to} 0)\propto \mbox{\large$\int$}\! \d p \:p^{2(\ell+1)}\delta(p^2\!/2{-}E) \propto E^{\ell+1/2},
\end{equation}
where $\ell$ is the angular momentum of the fragment pair. 
Following the intuition that ZEPE originates in the combination of the Wigner law and a Gaussian energy distribution induced by the short laser pulse, located at the binding energy $-E_{\rm EA}$ due to the two-photon zero-energy process, the detachment probability should be given by
\begin{subequations}\label{eq:ZEPE-spectrum}\begin{align}
 P_{\rm ZEPE}^{\ell,\beta}(E) &=P_*(E)\,s_{\ell,\beta}(E/E_{\rm EA}),
 \tag{\ref{eq:ZEPE-spectrum}}
\\
 P_*(E) & = [1+E/E_*]^{-1} \label{eq:back}
\\
 s_{\ell,\beta}(x) & = \beta^4x^{\ell+\frac{1}{2}}\exp(-\beta^2[x{+}1]^2).
\end{align}\end{subequations} 
The universal shape $s_{\ell,\beta}(x)$ of the zero-energy photo-electron spectrum is fully determined by the angular momentum $\ell$ of the photo-electron and $\beta\,{=}\,E_{\rm EA}/\Delta E$ the ratio of electron affinity $E_{\rm EA}$ to spectral pulse width $\Delta E$.
From the spectral representation of the pulse and the fact that ZEPE is a two-photon process follows $\Delta E\,{=}\,2/T$, which  is confirmed by 2nd-order perturbation theory discussed below.
Furthermore, $P_*(E)$ takes care of the slowly-varying background, being different for each ion.

\begin{figure*}[t]
 \centering
 \includegraphics[width=\textwidth]{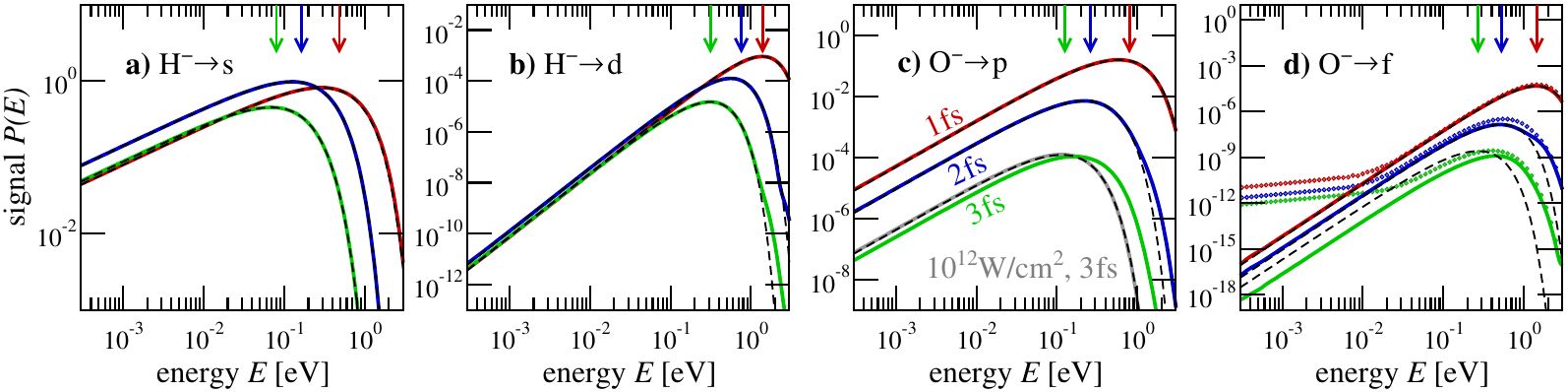}
 \caption{Photo-electron spectra for $H^-$ (a,b) as in Fig.\,\ref{fig:ZEPE-linear} but also for the two final channels of O$^-$ (c,d) in double-logarithmic scale.
 Note the different vertical scale for each panel.
 The gray line in panel (c) is for a lower intensity of $I{=}10^{12}$W/cm$^2$ and therefore scaled by a factor of $10^4$.
 The dotted lines in panel (d) are computed with a basis of states up to $E_{\rm cut}{=}30$\,eV which would allow to describe 3-photon processes, the solid curves are for $E_{\rm cut}{=}3$\,keV.
 Arrows mark the maxima according to Eq.\,\eqref{eq:emax}.
 }
 \label{fig:ZEPE-logarithmic}
\end{figure*}%
The dashed lines in Fig.\,\ref{fig:ZEPE-linear} show $P^{\ell,\beta}_{\rm ZEPE}(E)$ according to Eq.\,\eqref{eq:ZEPE-spectrum}.
The only fit parameter $E_\ast$ assumes the values
$E_{\ast}{=}0.896$\,eV and $E_{\ast}{=}0.684$\,eV for the ``s'' and ``p'' photo-electrons of hydrogen and oxygen, respectively, while $E_{\ast}{=}13.4$\,eV and $E_{\ast}{>}1000$\,eV for the electrons detached with angular momentum ``d'' and ``f'' reveal that the background is nearly constant in these cases over the energy interval considered. 
Obviously, Eq.\,\eqref{eq:ZEPE-spectrum} represents the numerical results very well, giving confidence in the interpretation and description of ZEPE provided. 
Deviations for O$^-$ at ``longer'' pulses can be attributed to higher-order effects occurring for intense pulses, since the agreement with the analytic description for a weaker pulse with $I{=}10^{12}$W/cm$^2$, as shown by the gray line in Fig.\,\ref{fig:ZEPE-logarithmic}c, is excellent.

Peculiar at second glance is the variation of the maximal detachment with the pulse length $T$ or $\Delta E$, respectively. Note, that the highest maximum is achieved in Fig.\ref{fig:ZEPE-linear} with $T_{\rm fwhm}{=}\,2$\,fs for the $\ell{=}0$ case, while for all other $\ell$ the maxima seem to increase monotonically. 
To elucidate their behavior systematically, we determine the maximum of $P_{\rm ZEPE}^{\ell,\beta}$ with respect to the two-dimensional parameter space $\{\beta, x\}$, i.\,e., the (scaled) pulse width $\beta\,{=}\, E_{\rm EA}/\Delta E$ and the (scaled) excess energy $x\,{=}\,E/E_{\rm EA}$.
The corresponding optimal parameters are given by analytical but lengthy expressions. 
Inserting $\beta(x)\,{=}\,\sqrt{2}/(1{+}x)$, the solution of $\partial\!P_{\rm ZEPE}^{\ell,\beta}/\partial\!\beta {=} 0$ into $\partial\!P_{\rm ZEPE}^{\ell,\beta}/\partial\!x {=} 0$ gives an analytical but lengthy expression for $\beta_{\rm Max}(x^*,\ell)$, i.\,e.\ the pulse width that gives the highest peak of the detachment probability.
Ignoring the background modification \eqref{eq:back} by taking the limit $x^*{\to}\infty$ one obtains with the
simpler expressions $\beta_{\rm Max}\,{=}\,(7{-}2\ell)/{4\sqrt{2}}$ and $x_{\rm Max}\,{=}\,(1{+}2\ell)/(7{-}2\ell)$ directly the optimal pulse duration $T_{\rm fwhm}{=}\sqrt{8\log2}\,\beta_{\rm Max}/E_{\rm EA}$.
They values are 2.55\,fs and 1.09\,fs for H$^-$ (s- and d-channel) and 0.92\,fs and 0.18\,fs for O$^-$ (p- and f-channel), consistent with the data shown in Fig.\,\ref{fig:ZEPE-logarithmic}.
Within the same approximation one can determine the location of the maxima of $P_{\rm ZEPE}^{\ell,\beta}$ for a given pulse durations $\beta$.
They read
\begin{equation}\label{eq:emax}
E_{\rm max} = \frac{E_{\rm EA}}{2}\big(\sqrt{1{+}(2\ell{+}1)/\beta^{2}} -1\big) 
\end{equation}
and are shown in Fig.\,\ref{fig:ZEPE-logarithmic} for the respective pulses with vertical arrows.
Although for $\ell{=}0$ (H$^-$) and $\ell{=}1$ (O$^-$) the background is relevant, since $E_*$ is in the energy range of interest, the values \eqref{eq:emax} explain the maxima there quite well.

Hence, the overview of ZEPE for H$^{-}$ and O$^{-}$ in Fig.\,\ref{fig:ZEPE-logarithmic}, highlighting the influence of different partial waves $\ell = 0,...,3$ of the fragments reveal that essentially all features discussed so far can be identified and are confirmed. For completeness we estimate 
the number of ZEPE electrons per shot one could detect in an experiment,
\begin{equation}
N_{\rm exp}=N_{\rm ion}
\big[I_{\rm exp}\big/10^{14}{\rm W/cm}^2\big]^{2}
P_T,
\end{equation}
where $N_{\rm ion}$ is the number of ions in the target volume, $I_{\rm exp}$ the experimental laser intensity and $P_T$ is given in Tab.\,\ref{tab:pt}.

\begin{figure}[t]
 \includegraphics[width=\columnwidth]{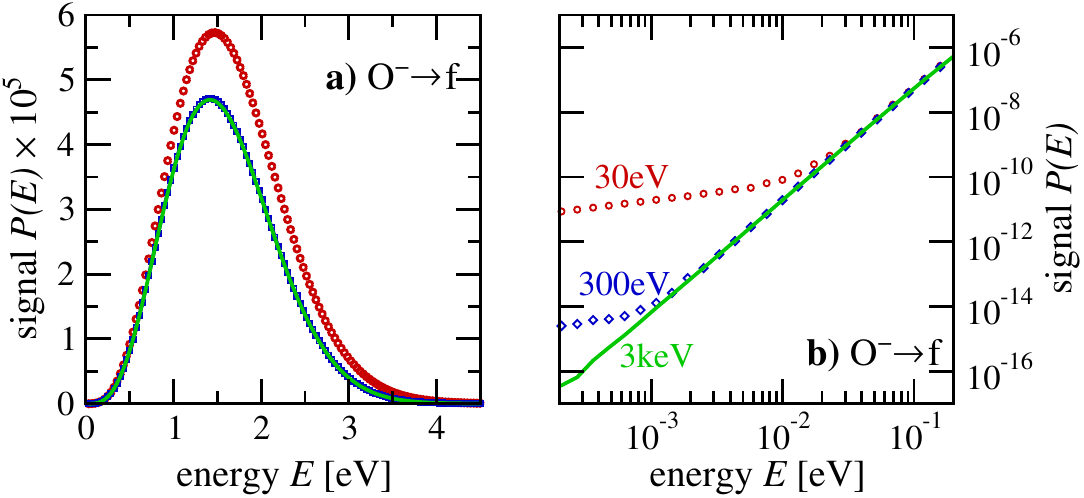}
 \caption{Photo-electron spectra in linear (left panel) and double-logarithmic (right) scale for the f-channel of O$^-$ from TDSE calculations ($\omega$ and $I $ as in Fig.\,\ref{fig:ZEPE-linear}, $T_{\rm fwhm}{=}1$\,fs) with different cut-off energies $E_{\rm cut}$, see text.}
 \label{fig:ZEPE-basis}
\end{figure}

\begin{table}[b]
\centering
 \caption{Total ionization probability $P_{T}$ for H$^-$ and O$^-$ for accessible channels and three pulse durations $T_{\rm fwhm}$.}
 \label{tab:pt}
\begin{tabular}{rcccccccc}
\hline\hline
&& H$^-{\to}\,$s && H$^-{\to}\,$d && O$^-{\to}\,$s && O$^-{\to}\,$d \\
$T_{\rm fwhm}$=1fs && $8.9{\times}10^{-4}$ && $1.5{\times}10^{-6}$ 
 && $1.8{\times}10^{-4}$ && $7.1{\times}10^{-8}$ \\
2fs && $4.4{\times}10^{-4}$ && $9.1{\times}10^{-8}$ 
 && $3.1{\times}10^{-6}$ && $9.7{\times}10^{-11}$ \\
3fs && $1.1{\times}10^{-4}$ && $6.4{\times}10^{-9}$ 
 && $3.9{\times}10^{-8}$ && $6.8{\times}10^{-13}$\\
\hline\hline
\end{tabular}
\end{table}%
The double-logarithmic scale in Fig.\,\ref{fig:ZEPE-logarithmic} was chosen to emphasize the threshold behavior and one can see that the quasi-analytical formula \eqref{eq:ZEPE-spectrum} performs in all cases
extremely well compared to the numerical TDSE calculations, particularly with regard to the large dynamic ranges considered. 
In fact, the TDSE results in panel \ref{fig:ZEPE-logarithmic}d contain a surprise: They appear not to be converged for the $\ell\,{=}\,3$ spectrum of oxygen.
Indeed, despite very small photo-electron energies of the order of 1\,meV, numerically one has to include continuum electrons up to 3\,keV to achieve convergence (see also Fig.\,\ref{fig:ZEPE-basis}).
An explanation is provided by 2nd-order perturbation theory for the two-photon zero-energy process comprising the two different events of absorption followed by emission ($\eta{=}``+"$), typically much stronger than emission followed by absorption ($\eta{=} ``-"$). The ionization amplitude to energy $E$ for these processes reads \cite{az23}
\begin{widetext}\begin{align}
a(E)
& = \sum_{k}d_{Ek}d_{k\,\rm EA}\int_{-\infty}^{+\infty}\!\!\!\d t\, A(t)\e{\i[E-E_k] t}
\int_{-\infty}^{t}\!\!\!\d t'\, A(t')\e{\i[E_{k}+E_{\rm EA}] t'}
\nonumber
\\
& = \frac{\pi}{8}T^{2}A_{0}{\!}^{2}
\sum_{k}d_{Ek}d_{k\,\rm EA}\,\sum_{\eta=\pm}\Big[\e{-[[\Delta^\eta_{k}+E_{\rm EA}]^{2}+[E-\Delta^\eta_{k}]^{2}]T^{2}\!/4}-\tfrac{2\i}{\sqrt{\pi}}\e{-[E+E_{\rm EA}]^{2}T^{2}\!/8}\,F\big([2\Delta^\eta_{k}{+}E_{\rm EA}{-}E]T/\sqrt{8}\big)\Big]
\label{eq:2nd-order}
\end{align}\end{widetext}
with the dipole matrix elements $d_{jk}\,{\equiv}\,\vec{e}_{z}\,{\cdot}\,\langle\phi_{j}|\vec{d}|\phi_{k}\rangle$ and the detunings $\Delta^\pm_k\,{\equiv}\,E_{k}{\mp}\omega$. 
We have not explicitly specified that $|\phi_0\rangle$, $|\phi_k\rangle$, $|\phi_E\rangle$ have different angular momenta $\ell$, dictated by selection rules in the dipole matrix elements,
and for simplicity we have assumed that all virtual states $|\phi_k\rangle$ are discrete in accordance with our numerical treatment of the continuum.
Finally, due to the necessary cutoff at $E_{\rm cut}$, the sum over virtual states $k$ is finite.
From the real part of \eqref{eq:2nd-order} one sees that continuum states, which are resonant with initial one-photon absorption $\Delta_k\,{=}\,0$, are dominant as well as the final energy equal to the initial energy, $E\,{=}\,{-}E_{\rm EA}$, as expected for the zero-energy photoelectric effect.

Coming back to the requirement of including very high energies $E_k$ in the calculation to reach convergence for small finite energy $E$, we note that the Dawson function $F$ \cite{note1} in the imaginary part of Eq.\,\eqref{eq:2nd-order} falls off very slowly $F(x)\,{\sim}\,1/(2x)$ for large $x$.
That implies that non-resonant states $E_k\gg E$ can contribute significantly.
Since the detachment probability ${\propto}\,(E/E_{\rm EA})^{\ell+1/2}$ is very small near threshold, it must be determined with high absolute accuracy which explains why unexpectedly high virtual energies need to be taken into account. Note, that for the very same reason the two-photon process of first {\em emitting} and then absorbing a photon, in standard situations never considered, has a non-negligible contribution.

In summary, we have discussed a universal two-photon process optimally observable in negative ions, where the weakly-bound electron interacts with a short laser pulse of spectral width comparable to the electronic binding energy.
Together with the typical Wigner power law of the electron-detachment cross section near threshold, a characteristic maximum forms at very-low electron energies, which does not dependent on the photon frequency, as accurate numerical calculations confirm.
They require the inclusion of surprisingly energetic electronic continuum states to converge, highlighting subtle features of this unconventional process which also surface in 2nd-order time-dependent perturbation theory. 
Following physical intuition, we have derived an analytical form of the spectrum. 
It describes the relative peak height and location as well as the dependence on pulse length and electron affinity of this peculiar zero-energy photoelectric effect very well which will facilitate its experimental realization.

S.\,A.\ acknowledges discussion with Jonathan Dubois in the early stage of this work.

\def\articletitle#1{\emph{#1}}


\begin{thebibliography}{10}
\bibitem{note1}
The Dawson function $F(z)$ \cite {da97} is related to the (complex) error
 function, $F(z)=\protect \frac {\protect \sqrt {\pi }}{2\protect \mathrm
 {i}}\protect \mathrm {e}^{{-z^{2}}}\protect \mbox {erf}(\protect \mathrm
 {i}z)$.
\end{thebibliography}
\end{document}